\documentclass[11pt,a4paper]{article}
\usepackage{float}
\usepackage{multirow}
\usepackage{graphics}
\usepackage[cp1252,utf8]{inputenc}
\usepackage{hyperref}
\usepackage{graphicx}
\usepackage{wrapfig}
\usepackage{amsmath}
\usepackage{amsfonts}
\usepackage{amssymb}
\usepackage{relsize} 
\usepackage[top=1in, bottom=2cm, left=1.5cm, right=1.5cm]{geometry}
\usepackage{authblk}
\usepackage{physics}
\NewDocumentCommand{\tens}{t_}
{%
	\IfBooleanTF{#1}
	{\tensop}
	{\otimes}%
}
\NewDocumentCommand{\tensop}{m}
{%
	\mathbin{\mathop{\otimes}\displaylimits_{#1}}%
}
\usepackage{changepage} 

\usepackage{afterpage}

\usepackage{placeins}

\usepackage{wrapfig}
\usepackage{cite} 
\usepackage[nottoc,notlof,notlot]{tocbibind} 

\title{\bf Holographic study of entanglement and complexity for mixed states}
\author[a]{\bf  Ashis Saha \thanks{ashisphys18@klyuniv.ac.in}}
\author[b]{\bf Sunandan Gangopadhyay \thanks{sunandan.gangopadhyay@bose.res.in}}

\affil[a]{\textit{Department of Physics, University of Kalyani, Kalyani 741235, India}}
\affil[b]{\textit{Department of Theoretical Sciences,
 S.N.~Bose National Centre for Basic Sciences,}
\textit{JD Block, Sector-III, Salt Lake, Kolkata 700106, India}}

\date{}

\begin{document}

\maketitle

\begin{abstract}
\noindent In this paper, we holographically quantify the entanglement and complexity for mixed states by following the prescription of purification. The bulk theory we consider in this work is a hyperscaling violating solution, characterized by two parameters, hyperscaling violating exponent $\theta$ and dynamical exponent $z$. This geometry is dual to a non-relativistic strongly coupled theory with hidden Fermi surfaces. We first compute the holographic analogy of entanglement of purification (EoP), denoted as the minimal area of the entanglement wedge cross section and observe the effects of $z$ and $\theta$. Then in order to probe the mixed state complexity we compute the mutual complexity for the BTZ black hole and the hyperscaling violating geometry by incorporating the holographic subregion complexity conjecture. We carry this out for two disjoint subsystems separated by a distance and also when the subsystems are adjacent with subsystems making up the full system. Furthermore, various aspects of holographic entanglement entropy such as entanglement Smarr relation, Fisher information metric and the butterfly velocity has also been discussed. 
\end{abstract}

\section{Introduction}
The gauge/gravity duality \cite{Maldacena:1997re,Witten:1998qj,Aharony:1999ti} has been employed to holographically compute quantum information theoretic quantitites and has thereby helped us to understand the bulk-boundary relations. Among various observables of quantum information theory, entanglement entropy (EE) has been the most fundamental thing to study as it measures the correlation between two subsystems for a pure state. EE has a very simple definition yet sometimes it is notoriously difficult to compute. However, the holographic computation of entanglement entropy which can be denoted as the Ryu-Takayanagi (RT) prescription, is a remarkably simple technique which relates the area of a codimension-2 static minimal surface with the entanglement entropy of a subsystem \cite{Ryu:2006bv,Ryu:2006ef,Nishioka:2009un}. The RT-prescription along with its modification for time-dependent scenario (HRT prescription \cite{Hubeny:2007xt}) has been playing a key role for holographic studies of information theoretic quantitites as the perturbative calculations which could not be done on the field theoretic side due to its strongly coupled nature, can now be performed in the bulk side since it is of weakly coupled nature.\\
Another important information theoretic quantity which has gained much attention recently is the computational complexity. The complexity of a quantum state represents the minimum number of simple operations which takes the unentangled product state to a target state \cite{watrous2008quantum,aaronson2016complexity,Hashimoto:2018bmb}. There are several proposals to compute complexity holographically. Recently, several interesting attempts has been made to define complexity in QFT \cite{Jefferson:2017sdb,Chapman:2017rqy,Khan:2018rzm,Doroudiani:2019llj}. In context of holographic computation, initially, it was suggested that the complexity of a state (measured in gates) is proportional to the volume of the Einstein-Rosen bridge (ERB) which connects two boundaries of an eternal black hole \cite{Susskind:2014moa,Susskind:2014rva}
\begin{eqnarray}\label{CV1}
C_V(t_L,t_R)=\frac{V_{ERB}(t_L,t_R)}{8\pi R G_{d+1}}
\end{eqnarray}
where $R$ is the AdS radius and $V_{ERB}(t_L,t_R)$ is the co-dimension one extremal volume of ERB which is bounded by the two spatial slices at times $t_L$ and $t_R$ of two CFTs that live on the two boundaries of the eternal black hole. Another conjecture states that complexity can be obtained from the bulk action evaluated on the Wheeler-DeWitt patch \cite{Brown:2015bva,Brown:2015lvg,Goto:2018iay}

\begin{eqnarray}\label{CA}
C_A = \frac{I_{WDW}}{\pi \hbar}~.
\end{eqnarray}
The above two conjectures depends on the whole state of the physical system at the boundary. In addition to these proposals, there is another conjecture which depends on the reduced state of the system. This states that the co-dimension one volume enclosed by the co-dimension two extremal RT surface is proportional to the complexity
\begin{eqnarray}\label{HSC}
C_V=\frac{V(\Gamma_A^{min})}{8\pi R G_{d+1}}~.
\end{eqnarray}
This proposal is known as the holographic subregion complexity (HSC) conjecture in the literature \cite{Alishahiha:2015rta,Carmi:2016wjl,Karar:2017org}. Recently, in \cite{Gangopadhyay:2020xox}, it was shown that there exists a relation between the universal pieces of HEE and HSC. Furthermore, the universal piece of HSC is proportional to the sphere free energy $F_{S^p}$ for even dimensional dual CFTs and proportional to the Weyl $a$-anomaly for odd dimensional dual CFTs.\\
In recent times, much attention is being paid to the study of entanglement entropy and complexity for mixed states. For the study of EE for mixed states, the entanglement of purification (EoP) \cite{Terhal_2002} and entanglement negativity $\mathcal{E}$ \cite{Vidal:2002zz} has been the promising candidates. In the subsequent analysis, our focus will be on the computation of EoP. Consider a density matrix $\rho_{AB}$ corresponding to mixed state in Hilbert space $\mathcal{H}$, where $\mathcal{H}=\mathcal{H}_A \tens \mathcal{H}_B$. Now the process of purification states that one can construct a pure state $\ket{\psi}$ from $\rho_{AB}$ by adding auxillary degrees of freedom to the Hilbert space $\mathcal{H}$
\begin{eqnarray}
\rho_{AB} = tr_{A^{\prime}B^{\prime}}\ket{\psi}\bra{\psi};~\psi \in \mathcal{H}_{AA^{\prime}BB^{\prime}}=\mathcal{H}_{AA^{\prime}} \tens \mathcal{H}_{BB^{\prime}}~.
\end{eqnarray}
Such states $\psi$ are called purifications of $\rho_{AB}$. It is to be noted that the process of purification is not unique and different procedures for purification for the same mixed state exists. In this set up, the definition of EoP ($E_P$) reads \cite{Terhal_2002}
\begin{eqnarray}\label{EoP}
E_P(\rho_{AB}) = \mathop{min}_{ tr_{A^{\prime}B^{\prime}}\ket{\psi}\bra{\psi} }S(\rho_{AA^{\prime}});~\rho_{AA^{\prime}} = tr_{BB^{\prime}}\ket{\psi}\bra{\psi}~.
\end{eqnarray}
In the above expression, the minimization is taken over any state $\psi$ satisfying the condition $\rho_{AB} = tr_{A^{\prime}B^{\prime}}\ket{\psi}\bra{\psi}$, where $A^{\prime}B^{\prime}$ are arbitrary. In this paper we will compute the holographic analogy of EoP, given by the minimal area of entanglement wedge cross section (EWCS) $E_W$ \cite{Takayanagi:2017knl}. However, there is no direct proof of $E_P=E_W$ duality conjecture yet and it is mainly based on the following properties of $E_P$ which are also satisfied by $E_W$ \cite{Terhal_2002,Takayanagi:2017knl}. These properties are as follows.
\begin{eqnarray}\label{prop}
&(i)&~E_P(\rho_{AB}) = S(\rho_{A})=S(\rho_{B});~\rho_{AB}^2=\rho_{AB}\nonumber\\
&(i)&~\frac{1}{2} I(A:B) \leq E_P(\rho_{AB}) \leq min\left[S(\rho_{A}),S(\rho_{B})\right] \nonumber~.
\end{eqnarray}
In the above properties, $I(A:B)=S(A)+S(B)-S(A\cup B)$ is the mutual information between two subsystems $A$ and $B$. Further, there exists a critical separation length $D_c$ between $A$ and $B$ beyond which there is no connected phase for any $l$. At  $D_c$, $E_W$ probes the phase transition of the RT surface $\Gamma_{AB}^{min}$ between connected and disconnected phases. The disconnected phase is characterised by the condition mutual information $I(A:B)=0$. Some recent very interesting observations in this direction can be found in \cite{Espindola:2018ozt,Agon:2018lwq,Umemoto:2018jpc,Hirai:2018jwy,Bao:2019wcf,Kusuki:2019evw,Jeong:2019xdr,Umemoto:2019jlz,Harper:2019lff,Jokela:2019ebz,BabaeiVelni:2019pkw,Ghodrati:2019hnn,Boruch:2020wbe,BabaeiVelni:2020wfl,Chakrabortty:2020ptb}. Further, recently several measures for mixed states dual to EWCS has been proposed. Some of them are odd entropy \cite{Tamaoka:2018ned}, reflected entropy \cite{Dutta:2019gen,Chu:2019etd} and logarithmic negativity \cite{MohammadiMozaffar:2017chk,Kudler-Flam:2018qjo}. On the other hand, recently the study of complexity for mixed states has gained appreciable amount of attention \cite{Alishahiha:2018lfv,Caceres:2018blh,Agon:2018zso,Caceres:2019pgf}. Similar to the case of EE for mixed state, the concept of `purification' is also being employed in this context \cite{Agon:2018zso,Camargo:2020yfv}. The purification complexity is defined as the minimal pure state complexity among all possible purifications available for a mixed state. Preparing a mixed state on some Hilbert space $\mathcal{H}$, starting from a reference (pure) state involves the extension of the Hilbert space $\mathcal{H}$ by introducing auxillary degrees of freedom \cite{Caceres:2018blh,Caceres:2019pgf}. In this set up, a quantity denoted as the mutual complexity $\Delta C$ has been prescribed in order to probe the concept of purifiaction complexity \cite{Alishahiha:2018lfv,Caceres:2018blh,Caceres:2019pgf,Agon:2018zso}. The mutual complexity $\Delta C$ satisfies the following definition 
\begin{eqnarray}
\Delta\mathcal{C} = \mathcal{C}(\rho_{A})+ +\mathcal{C}(\rho_{B})-\mathcal{C}(\rho_{A\cup B})~.
\end{eqnarray}
In this paper, we will incorporate the HSC conjecture in order to compute the complexities $\mathcal{C}(\rho_{A})$, $\mathcal{C}(\rho_{B})$ and $\mathcal{C}(\rho_{A\cup B})$. We compute $\Delta\mathcal{C}$ in two different set ups. In one set up, we consider two disjoint subsystems $A$ and $B$ of width $l$ on the boundary Cauchy slice $\sigma$, separated by a distance $x$. We then compute the mutual complexity between these two subregions. The other set up, we consider that the boundary Cauchy slice $\sigma$ is a collection of two adjacent subsystems $A$ and $B$ of width $l$ with $A \cap B =0$ (zero overlap) and $A^c=B$. In this set up we compute the mutual complexity between a subregion $A$ and the full system $A \cup A^c$.\\

The paper is organized as follows. In Section \ref{sec2}, we briefly discuss the aspects of the bulk theory which in this case is a hyperscaling violating geometry. We then consider a single strip-like subsystem and holographically compute the EE in Section \ref{sec3}. We also make comments on the thermodynamical aspects of the computed HEE by computing the entanglement Smarr relation satisfied by the HEE. Furthermore, we holographically compute the relative entropy in order to obtain the Fisher information metric. In Section \ref{sec4}, we consider two strip-like subsystems and holographically compute the EoP by using the $E_P=E_W$ conjecture. We briefly study the temperature dependent behaviour of EWCS along with the effects of $z$ and $\theta$ on the $E_W$. The Butterfly velocity $v_B$ corresponding to the hyperscaling violating geometry is computed in \ref{Bvel}. We then compute the HSC corresponding to a single strip-like subsystem in Section \ref{sec5}. In Section \ref{sec6}, we holographically compute the mutual complexity $\Delta C$ by incorporating the HSC conjecture for the BTZ black hole and the hyperscaling violating geometry. We consider two different set ups to study the mutual complexity. We then conclude in Section \ref{sec7}. We also have an Appendix in the paper.

\section{Bulk theory: Hyperscaling violating geometry}\label{sec2}
We shall start our analysis with a bulk hyperscaling violating spacetime geometry. The solution corresponds to the following effective action of Einstein-Maxwell-scalar theory \cite{Ogawa:2011bz,Huijse:2011ef}
\begin{eqnarray}\label{1}
S_{bulk} = \frac{1}{16\pi G}\int d^{d+1}x \sqrt{-g} \left[\left(R-2\Lambda\right)-W(\phi)F_{\mu\nu}F^{\mu\nu}-\frac{1}{2}\partial_{\mu}\phi\partial^{\mu}\phi-V(\phi)\right]
\end{eqnarray}
where $F_{\mu\nu}$ is the Faraday tensor associated with the gauge field $A_{\mu}$, $\phi$ is the scalar field associated with the potential $V(\phi)$ and $W(\phi)$ is the coupling. Extremization of this action leads to the following black hole solution \cite{Huijse:2011ef}\footnote{Note that this geometry is different in form than the one considered in \cite{BabaeiVelni:2019pkw}.}
\begin{eqnarray}\label{2}
ds^2= \frac{R^2}{r^2}\left[-\frac{f(r)}{ r^{\frac{2(d-1)(z-1)}{(d-\theta-1)}}}dt^2 + r^{\frac{2\theta}{d-\theta-1}}\frac{dr^2}{f(r)}+\sum_{i=1}^{d-1}dx_i^2\right]~.
\end{eqnarray}	
The lapse function $f(r)$ has the form $f(r)= 1-\left(\frac{r}{r_h}\right)^{(d-1)(1+\frac{z}{d-\theta -1})}$ where $r_H$ is the event horizon of the black hole. The Hawking temperature of black hole is obtained to be 
\begin{eqnarray}\label{12}
T_H =\frac{(d-1)(z+ d-\theta -1)}{4\pi (d-\theta -1)}\frac{1}{r_{h}^{z(d-1)/(d-\theta -1)}}~.
\end{eqnarray}
The above mentioned metric is holographic dual to a $d$-dimensional non-relativistic strongly coupled theory with Fermi surfaces. The metric is associated with two independent exponents $z$ and $\theta$. The presence of these two exponents leads to the following scale transformations 
\begin{eqnarray}\label{3}
x_i &\rightarrow& \xi~x_i\nonumber\\
t &\rightarrow& \xi^z~t\nonumber\\
ds &\rightarrow& \xi^{\frac{\theta}{d-1}}~ds\nonumber~.
\end{eqnarray}
This non-trivial scale transformation of the proper spacetime interval $ds$ is quite different from the usual AdS/CFT picture. The non-invariance of $ds$ in the bulk theory implies violations of hyperscaling in the boundary theory. Keeping this in mind, $\theta$ is identified as the hyperscaling violation exponent and $z$ is identified as the dynamical exponent.  In the limit $z=1$, $\theta=0$, we recover the SAdS$_{d+1}$ solution which is dual to a relativistic CFT in $d$-dimensions and in the limit $z\neq1$, $\theta=0$, we obtain the `Lifshitz solutions'.\\
The two independent exponents $z$ and $\theta$ satisfy the following inequalities
\begin{eqnarray}\label{11}
\theta \leq d-2,~z \geq 1+\frac{\theta}{d-1}~.
\end{eqnarray}
The `equalities' of the above mentioned relations holds only for gauge theories of non-Fermi liquid states in $d=3$ \cite{PhysRevB.82.075127}. In this case, $\theta =1$ and $z=3/2$. For general $\theta=d-2$, logarithmic violation of the `Area law' of entanglement entropy \cite{Srednicki:1993im} is observed. This in turn means for $\theta = d-2$, the bulk theory holographically describes a strongly coupled dual theory with hidden Fermi surfaces. Some studies of information theoretic quantities for the above mentioned hyperscaling violating geometries can be found in \cite{MohammadiMozaffar:2017nri,Alishahiha:2018tep}.
\section{Holographic entanglement Smarr relation}\label{sec3}
To begin our analysis, we consider our subsystem $A$ to be a strip of volume $V_{sub}=L^{d-2}l$, where $-\frac{l}{2}<x_1<\frac{l}{2}$ and $-\frac{L}{2}<x_{2,3,..,d-1}<\frac{L}{2}$. The amount of Hawking entropy captured by the above mentioned volume reads
\begin{eqnarray}\label{10}
	S_{BH}=\frac{L^{d-2} l}{4G_{d+1} r_{h}^{d-1}}~.
\end{eqnarray}
It is to be noted that the thermal entropy of the dual field theory is related with the temperature\footnote{The thermal entropy of the dual field theory is basically the Hawking entropy of the black hole given in eq.(\ref{12}).} as $S_{th}\propto T^{\frac{d-1-\theta}{z}}$. For $\theta=d-2$, it reads $S_{th}\propto T^{\frac{1}{z}}$. This result is observed for compressible states with fermionic excitations.

We parametrize the co-dimension one static minimal surface as $x_1 = x_1(r)$ which leads to the following area of the extremal surface $\Gamma_A^{min}$
\begin{eqnarray}\label{4}
	A(\Gamma_A^{min}) = 2R^{d-1}L^{d-2}r_t^{\left(\frac{p}{p-\theta}\right)-p}\sum_{n=1}^{\infty}\frac{1}{\sqrt{\pi}}\frac{\Gamma(n+\frac{1}{2})}{\Gamma(n+1)}\alpha^{np\left(1+\frac{z}{p-\theta}\right)}\int_{0}^{1}du\frac{u^{\frac{\theta}{p-\theta}-p+np\left(1+\frac{z}{p-\theta}\right)}}{\sqrt{1-u^{2p}}};~u=\frac{r}{r_t},~\alpha=\frac{r_t}{r_h}
\end{eqnarray}
where $r_t$ is the turning point and $p=(d-1)$ which we have introduced for the sake of simplicity. By substituting the area functional (given in eq.(\ref{4})) in the RT formula, we obtain the HEE \cite{Ryu:2006bv}
\begin{eqnarray}\label{5}
S_{E} &=& \frac{A(\Gamma_A^{min})}{4G_{d+1}}\nonumber\\
&=& \frac{2L^{d-2}}{4G_{d+1}\left(\frac{p}{p-\theta}-p\right)} \left(\frac{1}{\epsilon}\right)^{p-\left(\frac{p}{p-\theta}\right)}+\frac{L^{d-2}r_t^{\left(\frac{p}{p-\theta}\right)-p}}{4G_{d+1}}\sum_{n=0}^{\infty}\frac{\Gamma(n+\frac{1}{2})}{\Gamma(n+1)}\frac{\alpha^{np\left(1+\frac{z}{p-\theta}\right)}}{p}\frac{\Gamma\left(\frac{\frac{p}{p-\theta}-p+np\left(1+\frac{z}{p-\theta}\right)}{2p}\right)}{\Gamma\left(\frac{\frac{p}{p-\theta}+np\left(1+\frac{z}{p-\theta}\right)}{2p}\right)}\nonumber~.\\
\end{eqnarray}
The relationship between the subsystem size $l$ and turning point $r_t$ reads (with the AdS radius $R=1$)
\begin{eqnarray}\label{6}
l = r_t^{\frac{p}{p-\theta}} \sum_{n=0}^{\infty}\frac{\Gamma(n+\frac{1}{2})}{\Gamma(n+1)}\frac{\alpha^{np\left(1+\frac{z}{p-\theta}\right)}}{p}\frac{\Gamma\left(\frac{\frac{p}{p-\theta}+p+np\left(1+\frac{z}{p-\theta}\right)}{2p}\right)}{\Gamma\left(\frac{\frac{p}{p-\theta}+2p+np\left(1+\frac{z}{p-\theta}\right)}{2p}\right)}~.
\end{eqnarray}
 We now proceed to probe the thermodynamical aspects of HEE. It can be observed from eq.(\ref{5}) that the exprssion of $S_{E}$ contains a subsystem independent divergent piece which we intend to get rid by defining a finite quantity. We call this finite quantity as the renormalized holographic entanglement entropy ($S_{REE}$). From the point of view of the dual field theory this divergence free quantity represents the change in entanglement entropy under an excitation. In order to obtain $S_{REE}$ holographically, firstly we need to compute the HEE corresponding to the asymptotic form ($r_h\rightarrow\infty$) of the hyperscaling violating black brane solution given in eq.(\ref{2}). This yields the following expression
\begin{eqnarray}\label{7}
S_{G}= \frac{2L^{d-2}}{4G_{d+1}\left(\frac{p}{p-\theta}-p\right)} \left(\frac{1}{\epsilon}\right)^{p-\left(\frac{p}{p-\theta}\right)} -\frac{2^{p-\theta}L^{d-2}}{4G_{d+1}\left(p-\frac{p}{p-\theta}\right)} \left(\frac{p-\theta}{p}\right)^{p-\theta-1}\frac{\pi^{\frac{p-\theta}{2}}}{l^{p-1-\theta}}\left(\frac{\Gamma\left(\frac{p+\frac{p}{p-\theta}}{2p}\right)}{\Gamma\left(\frac{\frac{p}{p-\theta}}{2p}\right)}\right)^{p-\theta}~.
\end{eqnarray}
We now subtract the above expression (which represents the HEE corresponding to the vacuum of the dual field theory) from $S_{E}$ (given in eq.(\ref{5})) in order to get a finite quantity $S_{REE}$. This can be formally represented as
\begin{eqnarray}\label{8}
S_{REE} = S_{E} - S_{G}~.
\end{eqnarray}
On the other hand, the internal energy $E$ of the black hole can be obtained by using the Hawking entropy (given in eq.(\ref{10})) and the Hawking temperature (given in eq.(\ref{12})). The computed expression of $E$ can be represented as
\begin{eqnarray}\label{13}
E = \left(\frac{p-\theta}{z+p-\theta}\right)S_{BH}T_H~.
\end{eqnarray}
This is nothing but the classical Smarr relation of BH thermodynamics. In \cite{Saha:2019ado}, it was shown that the quantity $S_{REE}$ and the internal energy $E$ satisfies a Smarr-like thermodynamic relation corresponding to a generalized temperature $T_g$. In this set up, this relation reads \cite{Saha:2020fon}
\begin{eqnarray}\label{9}
E= \left(\frac{p-\theta}{z+p-\theta}\right)S_{REE}T_g~.
\end{eqnarray}
It is remarkable to observe that the relation given in eq.(\ref{9}) has a striking similarity with the classical Smarr relation of BH thermodynamics, given in eq.(\ref{13}). In the limit $r_t \rightarrow r_h$, the leading term of the generalized temperature $T_g$ produces the exact Hawking temperature $T_H$  whereas in the limit $\frac{r_t}{r_h}\ll1$, the leading term of $T_g$ reads \cite{Saha:2020fon}
\begin{eqnarray}\label{14}
\frac{1}{T_g} = \Delta_1 l^z
\end{eqnarray} 
where the detailed expression of $\Delta_1$ reads
\begin{eqnarray}
\Delta_1 = \frac{2\pi^{3/2}}{p}\left(\frac{1}{\frac{p}{p-\theta}-p+p\left(1+\frac{z}{p-\theta}\right)}\right)\left(\frac{p}{\sqrt{\pi}}\frac{\Gamma\left(\frac{p+\frac{p}{p-\theta}}{2p}\right)}{\Gamma\left(\frac{2p+\frac{p}{p-\theta}}{2p}\right)}\right)^{1+z}\left(\frac{\Gamma\left(\frac{p+\frac{p}{p-\theta}+p(1+\frac{z}{p-\theta})}{2p}\right)}{\Gamma\left(\frac{2p+\frac{p}{p-\theta}+p(1+\frac{z}{p-\theta})}{2p}\right)}\right)\nonumber~.
\end{eqnarray}
From eq.(\ref{14}) it can be observed that in the UV limit, $T_g$ shows the similar behaviour as entanglement temperature $T_{ent}$ (proportional to the inverse of subsystem size $l$) \cite{Bhattacharya:2012mi}. 
\subsection{Relative entropy and the Fisher information metric}
We now proceed to compute the Fisher information metric for the hyperscaling violating geometry using the holographic proposal. The Fisher information metric measures the distance between two quantum states and is given by \cite{Banerjee:2017qti}
\begin{eqnarray}\label{29}
G_{F,\lambda\lambda} = \langle\delta\rho~\delta\rho\rangle^{(\sigma)}_{\lambda\lambda}=\frac{1}{2}~Tr\left(\delta\rho \frac{d}{d(\delta\lambda)}\log(\sigma+\delta\lambda\delta\rho)\vert_{\delta\lambda=0}\right)
\end{eqnarray}
where $\sigma$ is the density matrix and $\delta\rho$ is a small deviation from the density matrix. On the other hand, there exists a relation between the Fisher information metric and the relative entropy $S_{rel}$ \cite{Lashkari:2015hha}. This reads
\begin{eqnarray}\label{30}
G_{F,mm} = \frac{\partial^2}{\partial m^2}S_{rel}(\rho_m\vert\vert\rho_0);~S_{rel}(\rho_m\vert\vert\rho_0)=\Delta\langle H_{\rho_0}\rangle-\Delta S~.
\end{eqnarray}
In the above expression, $\Delta S$ is the change in the entanglement entropy from vacuum state, $\Delta\langle H_{\rho_0}\rangle$ is the change in the modular Hamiltonian and $m$ is the perturbation parameter. In this set up, we holographically compute the relative entropy $S_{rel}(\rho_m\vert\vert\rho_0)$. We consider the background is slightly perturbed from pure hyperscaling violating spacetime while the subsystem volume $L^{d-2}l$ is fixed. Then the inverse of the lapse function $f(r)$ (given in eq.(\ref{2})) can be expressed as
\begin{eqnarray}\label{31}
\frac{1}{f(r)}=1+mr^{p\left(1+\frac{z}{p-\theta}\right)}+m^2r^{2p\left(1+\frac{z}{p-\theta}\right)}
\end{eqnarray}
where $m=\left(\frac{1}{r_H}\right)^{p\left(1+\frac{z}{p-\theta}\right)}$ is the holographic perturbation parameter. Since we consider a perturbation to the background geometry and also consider that the subsystem size $l$ has not changed, we can express the turning point in the following perturbed form
\begin{eqnarray}\label{32}
r_t = r_t^{(0)} + m r_t^{(1)} + m^2 r_t^{(2)}
\end{eqnarray}
where $r_t^{(0)}$ is the turning point for the pure hyperscaling violating geometry and $r_t^{(1)}$, $r_t^{(2)}$ are the first and second order corrections to the turning point. We now write down the subsystem length $l$ upto second order in perturbation as
\begin{eqnarray}\label{33}
\frac{l}{r_t^{\frac{p}{p-\theta}}} = a_0 + m a_1r_t^{p\left(1+\frac{z}{p-\theta}\right)}+m^2a_2r_t^{2p\left(1+\frac{z}{p-\theta}\right)}
\end{eqnarray}
where
\begin{eqnarray}
a_0= \frac{\sqrt{\pi}}{p}\frac{\Gamma\left(\frac{\frac{p}{p-\theta}+p}{2p}\right)}{\Gamma\left(\frac{\frac{p}{p-\theta}}{2p}\right)},~a_1=\frac{\sqrt{\pi}}{2p}\frac{\Gamma\left(\frac{\frac{p}{p-\theta}+p+p\left(1+\frac{z}{p-\theta}\right)}{2p}\right)}{\Gamma\left(\frac{\frac{p}{p-\theta}+p\left(1+\frac{z}{p-\theta}\right)}{2p}\right)},~a_2=\frac{3\sqrt{\pi}}{8p}\frac{\Gamma\left(\frac{\frac{p}{p-\theta}+p+2p\left(1+\frac{z}{p-\theta}\right)}{2p}\right)}{\Gamma\left(\frac{\frac{p}{p-\theta}+2p\left(1+\frac{z}{p-\theta}\right)}{2p}\right)}\nonumber~.
\end{eqnarray}
Using eq.(\ref{32}) in eq.(\ref{33}) and keeping in mind the consideration that $l$ has not changed, we obtain the forms of $r_t^{(0)}$, $r_t^{(1)}$ and $r_t^{(2)}$
\begin{eqnarray}\label{34}
r_t^{(0)} = \left(\frac{l}{a_0}\right)^{\frac{p-\theta}{p}},~r_t^{(1)}=-\left(\frac{p-\theta}{p}\right)\left(\frac{a_1}{a_0}\right)\left(r_t^{(0)}\right)^{1+p\left(1+\frac{z}{p-\theta}\right)},~ r_t^{(2)}= \xi \left(r_t^{(0)}\right)^{1+2p\left(1+\frac{z}{p-\theta}\right)} 
\end{eqnarray}   
where
\begin{eqnarray}
	\xi = \left(\frac{p-\theta}{p}\right)\left[\left(\frac{2p-\theta}{2p}\right)\left(\frac{a_1}{a_0}\right)^2+(p-\theta)\left(1+\frac{z}{p-\theta}\right)\left(\frac{a_1}{a_0}\right)^2-\left(\frac{a_2}{a_0}\right)\right]~.
\end{eqnarray}
On a similar note, the expression for area of the static minimal surface upto second order in perturbation parameter $m$ can be obtained from eq.(\ref{4}). We then use eq.(\ref{32}) to recast the expression for the area of the minimal surface in the form
\begin{eqnarray}\label{35}
A(\Gamma_A^{min}) = A(\Gamma_A^{min})^{(0)}+mA(\Gamma_A^{min})^{(1)}+m^2 A(\Gamma_A^{min})^{(2)}~.
\end{eqnarray}
It has been observed that at first order in $m$, $S_{rel}$ vanishes \cite{Lashkari:2015hha} and in second order in $m$ it reads $S_{rel}=-\Delta S$. In this set up, it yields
\begin{eqnarray}\label{36}
S_{rel} = - m^2\frac{A(\Gamma_A^{min})^{(2)}}{4G_{d+1}}=m^2\frac{L^{d-2}}{4G_{d+1}}\Delta_2 \left(\frac{l}{a_0}\right)^{1+2z+(p-\theta)}
\end{eqnarray} 
where
\begin{eqnarray}
\Delta_2= 2p\left(\frac{p-\theta}{p}\right)\left(\frac{a_1^2}{a_0}\right)+p\left(p-\frac{\theta}{p-\theta}\right)\left(\frac{p-\theta}{p}\right)^2\left(\frac{a_1^2}{a_0}\right)-\left(\frac{2pa_2}{2p(1+\frac{z}{p-\theta})-p+\frac{p}{p-\theta}}\right)-2pa_0\xi\nonumber~.
\end{eqnarray}  
 By substituting the above expression in eq.(\ref{30}), the Fisher information is obtained to be
\begin{eqnarray}\label{37}
	G_{F,mm}= \frac{L^{d-2}}{2G_{d+1}}\Delta_2 \left(\frac{l}{a_0}\right)^{1+2z+(p-\theta)}\propto l^{d+2z-\theta}~.
\end{eqnarray}

\noindent In the limit $z=1$ and $\theta=0$, the above equation reads $G_{F,mm}\propto l^{d+2}$ which agrees with the result obtained in \cite{Karar:2019bwy}. The Fisher information corresponding to the Lifshitz type solutions can be found in \cite{Karar:2020cvz}.


\section{Entanglement wedge cross-section and the $E_P = E_W$ duality}\label{sec4}
We now proceed to compute the holographic entanglement of purification by considering two subsystems, namely, $A$ and $B$ of length $l$ on the boundary $\partial M$. From the bulk point of view, $\partial M$ is the boundary of a canonical time-slice $M$ made in the static gravity dual. Furthermore, $A$ and $B$ are separated by a distance $D$ so that the subsystems does not have an overlap of non-zero size ($A\cap B =0$). Following the RT prescription, we denote $\Gamma_A^{min}$, $\Gamma_B^{min}$ and $\Gamma_{AB}^{min}$ as the static minimal surfaces corresponding to $A$, $B$ and $AB$ respectively. In this set up, the domain of entanglement wedge $M_{AB}$ is the region in the bulk with the following boundary 
\begin{eqnarray}\label{16}
\partial M_{AB} = A \cup B \cup \Gamma_{AB}^{min}~.
\end{eqnarray}
It is also to be noted that if the separation $D$ is effectively large then the codimension-0 bulk region $M_{AB}$ will be disconnected. We now divide $\Gamma_{AB}^{min}$ into two parts
\begin{eqnarray}\label{17}
\Gamma_{AB}^{min} = \Gamma_{AB}^{A} \cup \Gamma_{AB}^{B}
\end{eqnarray}
such that the boundary $\partial M_{AB}$ of the canonical time-slice of the full spacetime $M_{AB}$ can be represented as
\begin{eqnarray}
\partial M_{AB} = \bar{\Gamma}_A \cup \bar{\Gamma}_B
\end{eqnarray}
where 
$\bar{\Gamma}_A = A \cup \Gamma_{AB}^{A}$ and $\bar{\Gamma}_B = B \cup \Gamma_{AB}^{B}$. In this set up, it is now possible to define the holographic entanglement entropies $S(\rho_{A \cup \Gamma_{AB}^{A}})$ and $S(\rho_{B \cup \Gamma_{AB}^{B}})$. These quantities can be computed by finding a static minimal surface $\Sigma^{min}_{AB}$ such that
\begin{eqnarray}
\partial \Sigma^{min}_{AB} = \partial \bar{\Gamma}_A =  \partial \bar{\Gamma}_B~.
\end{eqnarray}
There can be infinite number possible choices for the spliting given in eq.(\ref{17}) and this in turn means there can be infinite number of choices for the surface $\Sigma^{min}_{AB}$. The entanglement wedge cross section (EWCS) is obtained by minimizing the area of $\Sigma^{min}_{AB}$ over all possible choices for $\Sigma^{min}_{AB}$. This can be formally written down as
\begin{eqnarray}\label{18}
	E_W(\rho_{AB}) = \mathop{min}_{\bar{\Gamma}_A \subset \partial M_{AB}}\left[\frac{A\left(\Sigma^{min}_{AB}\right)}{4G_{d+1}}\right]~.
\end{eqnarray}
We now proceed to compute $E_W$ for the holographic dual considered in this paper. As we have mentioned earlier, EWCS is the surface with minimal area which splits the entanglement wedge into two domains corresponding to $A$ and $B$. This can be identified as a vertical, constant $x$ hypersurface. The time induced metric on this constant $x$ hypersurface reads
\begin{eqnarray}
	ds_{ind}^2= \frac{R^2}{r^2}\left[ r^{\frac{2\theta}{p-\theta}}\frac{dr^2}{f(r)}+\sum_{i=1}^{d-2}dx_i^2\right]~.
\end{eqnarray}
By using this above mentioned induced metric, the EWCS is obtained to be
\begin{eqnarray}\label{15}
\begin{aligned}
E_W =&	\frac{L^{d-2}}{4G_{d+1}}\int_{r_t(D)}^{r_t(2l+D)}\frac{dr}{r^{d-1}\sqrt{f(r)}}\nonumber\\
=& \frac{L^{d-2}}{4G_{d+1}}\sum_{n=0}^{\infty}\frac{1}{\sqrt{\pi}}\frac{\Gamma(n+\frac{1}{2})}{\Gamma(n+1)}\left[\frac{r_t(2l+D)^{np(1+\frac{z}{p-\theta})-p+1}-r_t(D)^{np(1+\frac{z}{p-\theta})-p+1}}{np(1+\frac{z}{p-\theta})-p+1}\right]\left(\frac{1}{r_h}\right)^{np(1+\frac{z}{p-\theta})}~.
\end{aligned}
\end{eqnarray}
As mentioned earlier, the above expression of $E_W$ always maintains the following bound
\begin{eqnarray}\label{bound}
	E_W \geq \frac{1}{2}I(A:B);~I(A:B)=S(A)+S(B)-S(A\cup B)
\end{eqnarray}
where $I(A:B)$ is the mutual information between two subsystems $A$ and $B$. On the other hand, in \cite{Yang:2018gfq} it was shown that there exists a critical separation between $A$ and $B$ beyond which there is no connected phase for any $l$. This in turn means that at the critical separation length $D_c$, $E_W$ probes the phase transition of the RT surface $\Gamma_{AB}^{min}$ between connected and disconnected phases. The disconnected phase is characterised by the fact that the mutual information $I(A:B)$ vanishes, which in this case reads
\begin{eqnarray}
2S(l)-S(D)-S(2l+D)=0~.
\end{eqnarray}
The above condition together with the bound given in eq.(\ref{bound}) leads to the critical separation length $D_c$ \cite{Ben-Ami:2014gsa}.\\

\noindent We now write down the expression for $E_W$ (given in eq.(\ref{15})) in terms of the parameters of the boundary theory. This we do for the small temperature case ($\frac{r_t(D)}{r_h}\ll \frac{r_t(2l+D)}{r_h} \ll 1 $) and for the high temperature case $(\frac{r_t(D)}{r_h}\ll 1, \frac{r_t(2l+D)}{r_h}\approx1)$.
\subsection{$E_W$ in the low temperature limit}
In the limit $\frac{r_t(D)}{r_h}\ll \frac{r_t(2l+D)}{r_h} \ll 1 $, it is reasonable to consider terms upto order $m$ (where $m= \frac{1}{r_h^{p(1+\frac{z}{p-\theta})}}$) in the expression for $E_W$ (given in eq.(\ref{15})). On the other hand for low temeprature considerations, it is possible to perturbatively solve eq.(\ref{6}) which leads to the following relationship between a subsystem size $l$ and its corresponding turning point $r_t$
\begin{eqnarray}\label{19}
	r_t(l) = l^{\frac{p-\theta}{p}}\left(\frac{p}{\sqrt{\pi}}\frac{\Gamma\left(\frac{2p+\frac{p}{p-\theta}}{2p}\right)}{\Gamma\left(\frac{p+\frac{p}{p-\theta}}{2p}\right)}\right)^{\frac{p-\theta}{p}}\times \left[1- \Delta_3 l^{(p-\theta)\left(1+\frac{z}{p-\theta}\right)}T^{\left(1+\frac{p-\theta}{z}\right)}
	\right]
\end{eqnarray}
where 
\begin{eqnarray}
\Delta_3 &=& \left(\frac{p-\theta}{2p}\right)\left(\frac{4\pi}{p(1+\frac{z}{p-\theta})}\right)^{1+\frac{p-\theta}{z}} \left(\frac{\Gamma\left(\frac{2p+\frac{p}{p-\theta}}{2p}\right)}{\Gamma\left(\frac{p+\frac{p}{p-\theta}}{2p}\right)}\right)\left(\frac{\Gamma\left(\frac{p+\frac{p}{p-\theta}+p(1+\frac{z}{p-\theta})}{2p}\right)}{\Gamma\left(\frac{2p+\frac{p}{p-\theta}+p(1+\frac{z}{p-\theta})}{2p}\right)}\right)\left(\frac{p}{\sqrt{\pi}}\frac{\Gamma\left(\frac{2p+\frac{p}{p-\theta}}{2p}\right)}{\Gamma\left(\frac{p+\frac{p}{p-\theta}}{2p}\right)}\right)^{(p-\theta)\left(1+\frac{z}{p-\theta}\right)}\nonumber~.
\end{eqnarray}
Now by using eq.(\ref{19}) for $r_t(2l+D)$ and $r_t(D)$, we obtain the expression for $E_W$ in the low temperature limit to be
\begin{eqnarray}\label{20}
	E_W = E_W^{T=0} - \frac{L^{d-2}}{4G_{d+1}}\Delta_4\left[(2l+D)^{\left(\frac{p-\theta}{p}\right)\left(1+\frac{pz}{p-\theta}\right)}-D^{{\left(\frac{p-\theta}{p}\right)\left(1+\frac{pz}{p-\theta}\right)}}\right]T^{1+\left(\frac{p-\theta}{z}\right)}+...
\end{eqnarray}
where the detailed expression for $\Delta_4$ is given in the Appendix. The first term in eq.(\ref{20}) is the EWCS at $T=0$. This reads
\begin{eqnarray}\label{21}
E_W^{T=0} = \frac{L^{d-2}}{4(p-1)G_{d+1}} \left(\frac{\sqrt{\pi}}{p}\frac{\Gamma\left(\frac{p+\frac{p}{p-\theta}}{2p}\right)}{\Gamma\left(\frac{2p+\frac{p}{p-\theta}}{2p}\right)}\right)^{(p-\theta)-\left(\frac{p-\theta}{p}\right)}\left[\left(\frac{1}{D}\right)^{(p-\theta)-\left(\frac{p-\theta}{p}\right)}-\left(\frac{1}{2l+D}\right)^{(p-\theta)-\left(\frac{p-\theta}{p}\right)}\right]~.\nonumber\\
\end{eqnarray}
It can be observed from eq.(\ref{20}) that the EWCS is a monotonically decreasing function of temperature $T$ (as $1+\frac{p-\theta}{z}>0$). We now compute the critical separation length $D_c$ at which $E_W$ probes the phase transition of the RT surface $\Gamma_{AB}^{min}$ between the connected and disconnected phases. This to be obtained from the condition
\begin{eqnarray}\label{MI}
 2S(l)-S(D)-S(2l+D)=0~.
\end{eqnarray}
The general expression for the HEE of a strip of length $l$ is given in eq.(\ref{5}). Now similar to the above computation, in the limit $\frac{r_t(D)}{r_h}\ll \frac{r_t(2l+D)}{r_h} \ll 1$, we consider terms upto $\mathcal{O}(m)$ in eq.(\ref{5}). By using this consideration, eq.(\ref{MI}) can be expressed as
\begin{eqnarray}
\beta_1\left(\frac{2}{l^{p-\theta-1}}-\frac{1}{D^{p-\theta-1}}-\frac{1}{(2l+D)^{p-\theta-1}}\right)+\frac{\beta_2}{r_h^{p(1+\frac{z}{p-\theta})}}\left(2l^{1+z}-D^{1+z}-(2l+D)^{1+z}\right)=0
\end{eqnarray}
where
\begin{eqnarray}
	\beta_1 &=& \frac{\sqrt{\pi}}{4p} \frac{\Gamma\left(\frac{\frac{p}{p-\theta}-p}{2p}\right)}{\Gamma\left(\frac{\frac{p}{p-\theta}}{2p}\right)}\left(\frac{\sqrt{\pi}}{p}\frac{\Gamma\left(p+\frac{\frac{p}{p-\theta}}{2p}\right)}{\Gamma\left(\frac{2p+\frac{p}{p-\theta}}{2p}\right)}\right)^{p-\theta-1}\nonumber\\
\beta_2&=&\Delta_3\left(p-\frac{p}{p-\theta}\right)\beta_1+\frac{\sqrt{\pi}}{8p}\left(\frac{p}{\sqrt{\pi}}\frac{\Gamma\left(\frac{p+\frac{p}{p-\theta}}{2p}\right)}{\Gamma\left(\frac{2p+\frac{p}{p-\theta}}{2p}\right)}\right)^{1+z}\left(\frac{\Gamma\left(\frac{p+\frac{p}{p-\theta}+p(1+\frac{z}{p-\theta})}{2p}\right)}{\Gamma\left(\frac{2p+\frac{p}{p-\theta}+p(1+\frac{z}{p-\theta})}{2p}\right)}\right)~.\nonumber	
\end{eqnarray}
By solving the above equation we can find out the critical separation length $D_c$ (where we substitute $\frac{D}{l}=k=constant$). It is worth mentioning that in the above computations of $E_W$ and $I(A:B)$, we have considered terms upto $\mathcal{O}(m)$ which is the leading order term for thermal correction. Similarly, one can incorporate next-to leading order terms or more to get a more accurate result.
\begin{figure}[!h]
	\begin{minipage}[t]{0.48\textwidth}
		\centering\includegraphics[width=\textwidth]{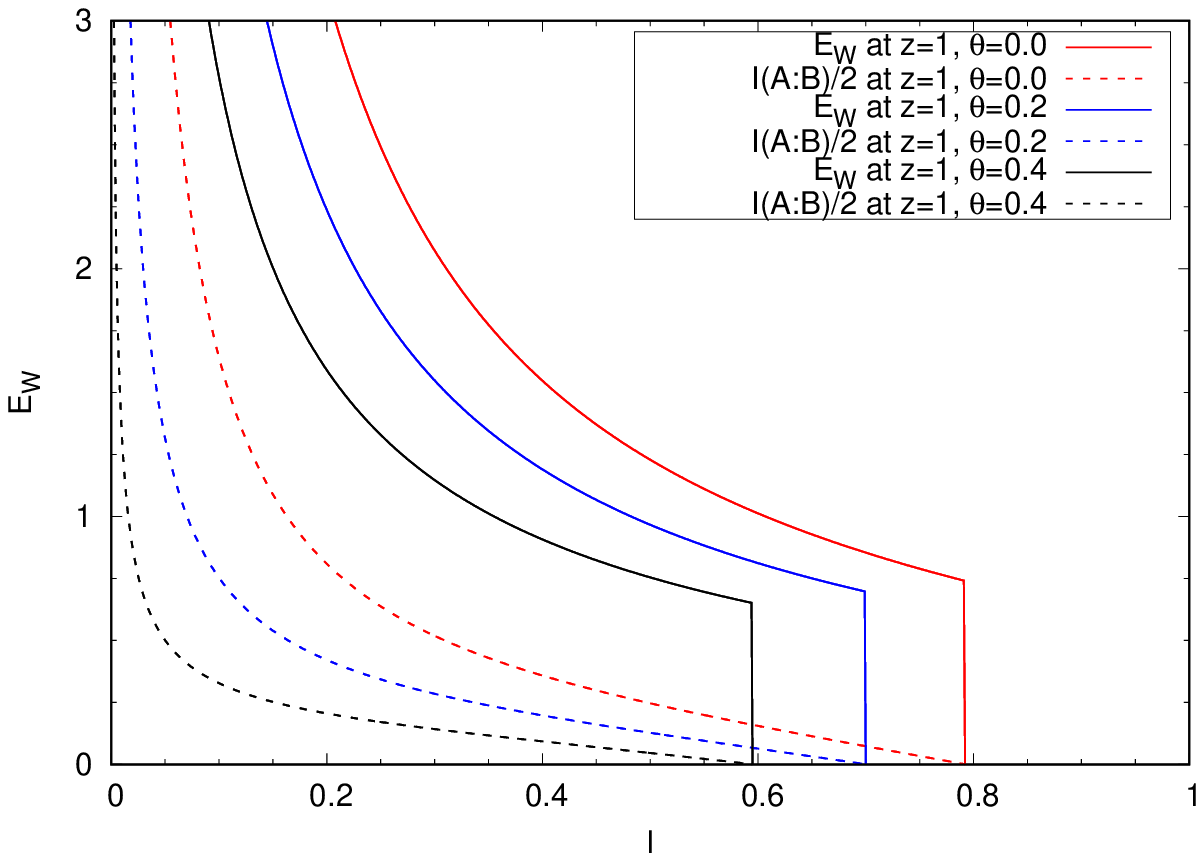}\\
		{\footnotesize Effect of hyperscaling violating exponent $\theta$ (we set $z=1$)}
	\end{minipage}\hfill
	\begin{minipage}[t]{0.48\textwidth}
		\centering\includegraphics[width=\textwidth]{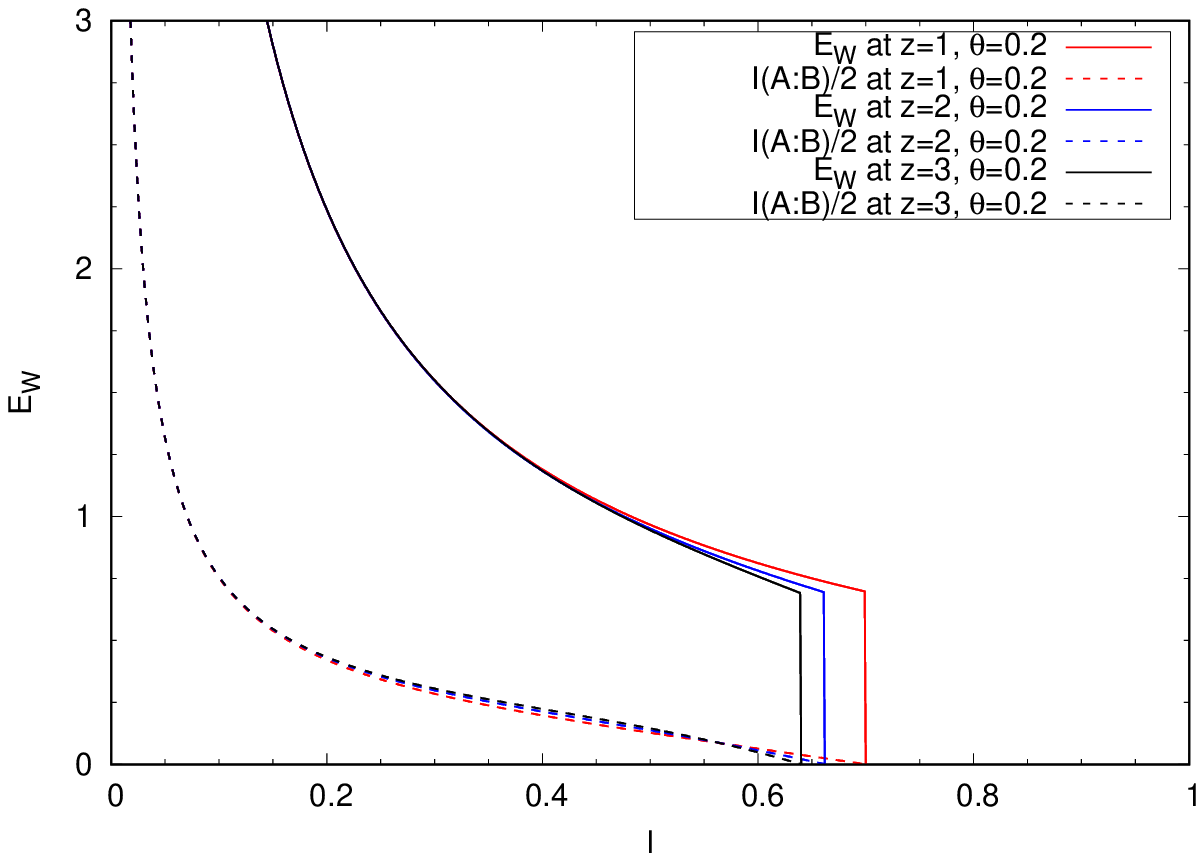}\\
		{\footnotesize Effect of dynamical exponent $z$ (we set $\theta=0.2$)}
	\end{minipage}
	\caption{Effects of $\theta$ and $z$ on $E_W$ and $I(A:B)$ at low temperature (with $d=3$, $k=0.4$, $L=1$ and $G_{d+1}=1$)}\label{fig2}
\end{figure}
\subsection{$E_W$ in the high temperature limit}
We now consider the limit $r_t(2l+D)\rightarrow r_h$ and $\frac{r_t(D)}{r_h}\ll 1$. This in turn means the static minimal surface associated with the turning point $r_t(2l+D)$ wraps a portion of the event horizon $r_h$. In the large $n$ limit, the infinite sum associated with the turning point $r_t(2l+D)$ goes as $\approx \frac{1}{\sqrt{\pi}}\left(\frac{1}{n}\right)^{3/2}\left(\frac{r_t(2l+D)}{r_h}\right)^{np(1+\frac{z}{p-\theta})}$ which means it is convergent ($\sum_{n=1}^{\infty} \frac{1}{n^{3/2}}=\xi(\frac{3}{2})$). Further, we are considering $\frac{r_t(D)}{r_h}\ll 1$ and in this limit it is reasonable to keep terms only upto order $m$ in the infinite sum associated with $r_t(D)$. In this set up, the expression for EWCS reads
\begin{eqnarray}\label{22}
E_W(T)=E_W^{T=0}+\frac{L^{d-2}}{4G_{d+1}}\Delta_4 D^{z+\left(\frac{p-\theta}{p}\right)}T^{1+\left(\frac{p-\theta}{z}\right)}-\frac{L^{d-2}}{4G_{d+1}}\Delta_5 T^{\left(\frac{p-\theta}{z}\right)-\left(\frac{p-\theta}{pz}\right)}+...
\end{eqnarray}
where the temperature independent term (EWCS at $T=0$) is being given by
\begin{eqnarray}\label{23}
E_W^{T=0}=\frac{L^{d-2}}{4(p-1)G_{d+1}D^{(\frac{p-\theta}{p})(p-1)}} \left(\frac{\sqrt{\pi}}{p}\frac{\Gamma\left(\frac{p+\frac{p}{p-\theta}}{2p}\right)}{\Gamma\left(\frac{2p+\frac{p}{p-\theta}}{2p}\right)}\right)^{(p-\theta)-\left(\frac{p-\theta}{p}\right)}~.
\end{eqnarray}
The expressions for $\Delta_4$ and $\Delta_5$ are given in the Appendix. Now we procced to compute the critical separation length $D_c$ in the high temperature configuration. In the limit $r_t \rightarrow r_h$, the computed result of $S_E$ (HEE of a strip like subsystem with length $l$), given in eq.(\ref{5}) can be rearranged in the following form\footnote{The first term of the expression is nothing but the thermal entropy of the boundary subsystem given in eq.(\ref{10}).}
\begin{eqnarray}
S_{E} = \frac{L^{d-2}l}{4G_{d+1}r_h^p}+\frac{L^{d-2}}{4G_{d+1}r_h^{p-\frac{p}{p-\theta}}}\sum_{n=0}^{\infty}P_n;~P_n =\left(\frac{1}{\frac{p}{p-\theta}-p+np(1+\frac{z}{p-\theta})}\right) \frac{\Gamma\left(n+\frac{1}{2}\right)}{\Gamma\left(n+1\right)} \frac{\Gamma\left(\frac{p+\frac{p}{p-\theta}+np(1+\frac{z}{p-\theta})}{2p}\right)}{\Gamma\left(\frac{2p+\frac{p}{p-\theta}+np(1+\frac{z}{p-\theta})}{2p}\right)}\nonumber~.\\
\end{eqnarray}
By using the above form of HEE, we can write down eq.(\ref{MI}) in the following form
\begin{eqnarray}
\frac{\sum_{n=0}^{\infty}P_n}{r_h^{p-\frac{p}{p-\theta}}}-\frac{D}{4r_h^{p-\frac{p}{p-\theta}}}-\frac{\beta_1}{D^{p-\theta-1}}-\frac{\beta_2}{r_h^{p(1+\frac{z}{p-\theta})}}D^{1+z}=0~.
\end{eqnarray}
\begin{figure}[!h]
	\begin{minipage}[t]{0.48\textwidth}
		\centering\includegraphics[width=\textwidth]{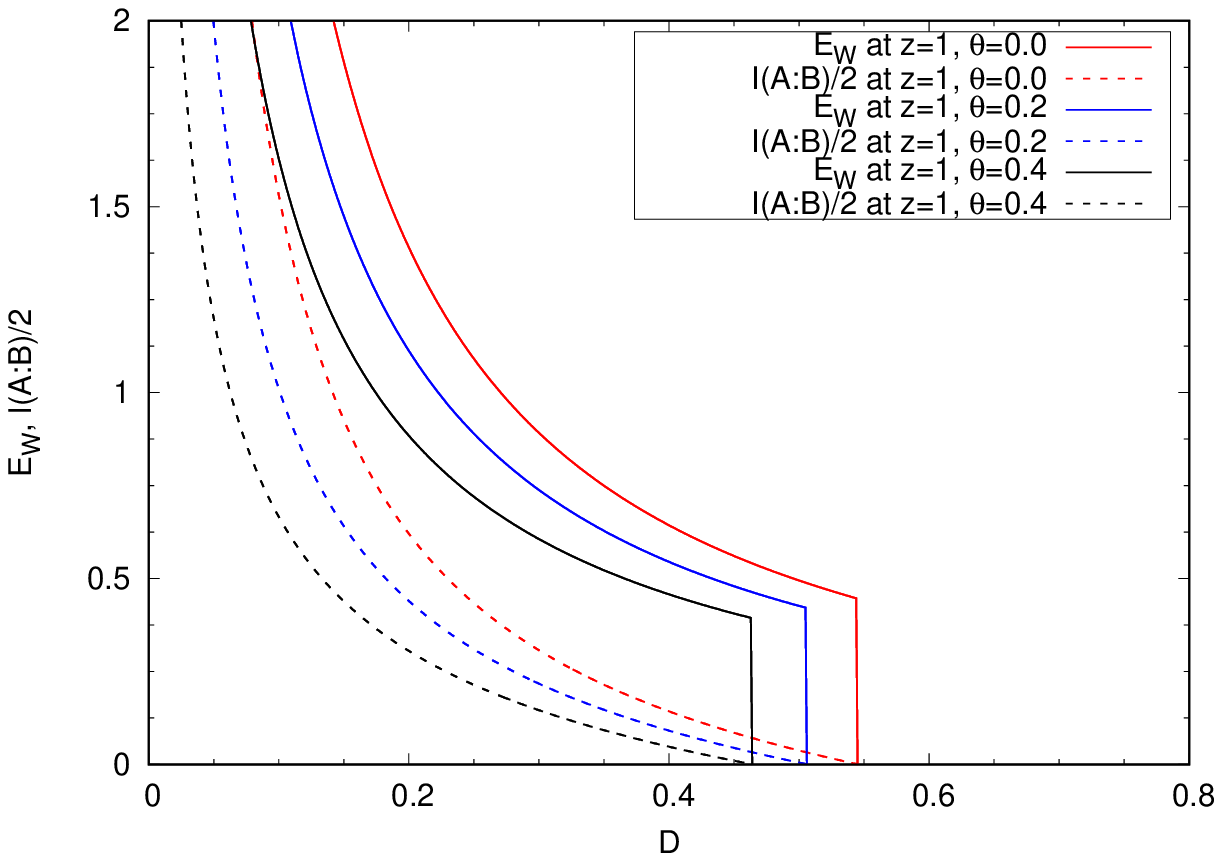}\\
		{\footnotesize Effect of hyperscaling violating exponent $\theta$ (we set $z=1$)}
	\end{minipage}\hfill
	\begin{minipage}[t]{0.48\textwidth}
		\centering\includegraphics[width=\textwidth]{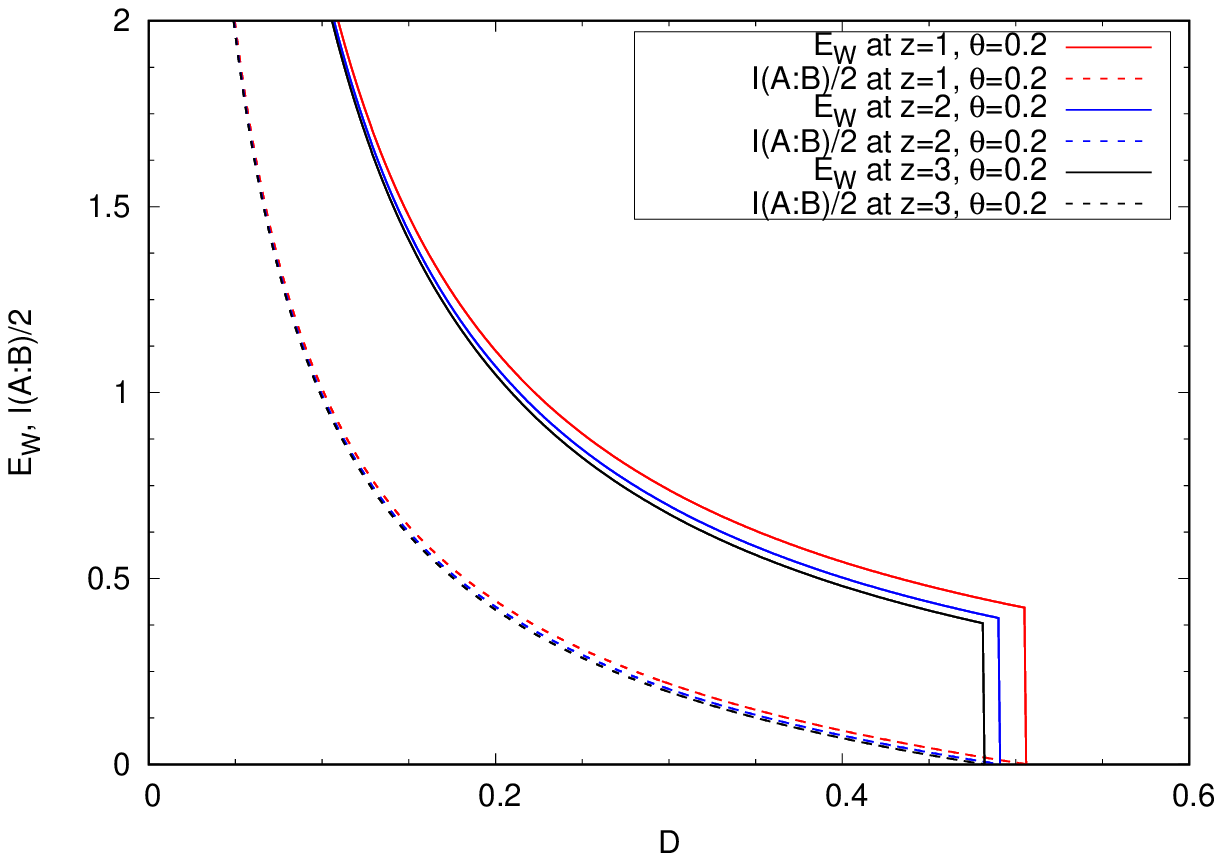}\\
		{\footnotesize Effect of dynamical exponent $z$ (we set $\theta=0.2$)}
	\end{minipage}
	\caption{Effects of $\theta$ and $z$ on $E_W$ and $I(A:B)$ at high temperature (with $d=3$, $L=1$ and $G_{d+1}=1$)}
	\label{fig3}
\end{figure}
In Fig(s).(\ref{fig2}) and (\ref{fig3}), we have graphically represented the effects of $z$ and $\theta$ on the EWCS and holographic mutual information (HMI) for both low and high temperature case respectively. For the low temperature case (Fig.(\ref{fig2})) we have chose the separation length $D$ between the subsystems to be $D = 0.4~l$. From the above plots it can be observed that the EWCS always maintains the bound $E_W > \frac{1}{2} I(A:B)$. The HMI continously decays and approaches zero at a particular critical separation length $D_c$. This critical separataion $D_c$ decreases with increasing $z$ and $\theta$. On the other hand $E_W$ shows a discontinous jump at $D_c$. Upto $D_c$, $E_W$ has a finite cross-section due to the connected phase whereas beyond this critical separation length $E_W$ vanishes due to the disconnected phase of the RT surface $\Gamma_{AB}^{min}$.
\section{Butterfly velocity}\label{Bvel}
In this section we shall discuss about information spreading in the dual field theory from the holographic point of view. In context of quantum many body physics, the study about the chaotic nature of the system (response of the system at a late time after a local perturbation at initial time) can be characterized by the following thermal average
\begin{eqnarray}\label{40}
C(x,t) = \langle\left[W(t,x),V(0)\right]^2\rangle_{\beta} 
\end{eqnarray}
where $V(0)$ is a generic operator acting at the origin at earlier time and $W(t,x)$ is a local operator acting at position $x$ at later time $t$. The butterfly effect is usually governed by such commutators. It probes the dependency of a late time measurement on the initial perturbation. The time at which the commutator grows to $\mathcal{O}(1)$, is known as the Scrambling time \cite{Hayden:2007cs}. The study of Butterfly effect in context of AdS/CFT naturally occurs as the black holes are observed to be the fastest Scramblers in the nature \cite{Sekino:2008he}. For large $N$ field theories, eq.(\ref{40}) grows as
\begin{eqnarray}
	C(x,t) = \frac{K}{N^2} \exp[\lambda_L\left(t-\frac{|x|}{v_B}\right)]+\mathcal{O}\left(\frac{1}{N^4}\right)
\end{eqnarray}
where $K$ is a constant, $\lambda_L$ is the Lyapunov exponent which probes the growth of chaos with time and $v_B$ is the Butterfly velocity. The Butterfly velocity $v_B$ probes the speed at which the local perturbation grows (the speed of the growth for chaos). This velocity composes a light cone for chaos and outside this light cone, the perturbation does not effects the system. Further, $v_B$ has the interpretation of effective Lieb-Robinson velocity (state-dependent) $v_{LR}$ for strongly coupled field theories \cite{Roberts:2016wdl}. The Lyapunov exponent safisfies a upper bound \cite{Maldacena:2015waa}
\begin{eqnarray}
\lambda_L \leq 2\pi T~.
\end{eqnarray}
 One can consider acting of a local operator (perturbation) on a thermal state of a CFT. At the initial stages, the information about the nature of the operator can be obatined by operating another local operator at position $x$. However, due to the Scrambling property, this information about the initial perturbation will spread out in a larger and larger region with time.\\
 In context of gauge/gravity duality, $v_B$ can be calculated by using the subregion duality \cite{Bousso:2012sj}. The static black hole in the bulk represents the initial thermal state in the dual field theory. One can add a local perturbation in this set up which eventually falls into event horizon. The time-like trajectories of the local perturbation in the bulk can be probed, by using the co-dimension two RT surfaces \cite{Shenker:2013pqa,Mezei:2016wfz}. In both of these scenario, there shall be a smallest subregion which will contain enough information (at later time $t$) about the local perturbation. It is aasumed that the bulk dual to this smallest subregion of the boundary is the entanglement wedge \cite{Czech:2012bh}. The Butterfly velocity represents the rate at which these subregions increases. The holographic computation requires only the near-horizon data about the dual gravitational solution. By following the approach given in \cite{Mezei:2016wfz}, a expression for the Butterfly velocity $v_B$ corresponding to a general black brane geometry has been computed in \cite{DiNunno:2021eyf}. For hyperscaling violating geometry (given in eq. (\ref{2})), the expression for $v_B$ is obtained to be
 \begin{eqnarray}\label{41}
 	v_B = \sqrt{\frac{1}{2}\left(1+\frac{z}{d-\theta-1}\right)} \left[\frac{4\pi}{(d-1)\left(1+\frac{z}{d-\theta-1}\right)}\right]^{1-\frac{1}{z}} T^{1-\frac{1}{z}} \propto T^{1-\frac{1}{z}}.
 \end{eqnarray}     
In the AdS limit $z\rightarrow 1, \theta\rightarrow 0$, it reduces to the well-known result $v_B = \sqrt{\frac{d}{2(d-1)}}$ \cite{Mezei:2016wfz}. It is known that $v_B$ is a model dependent parameter which in this case captures the collective effects of $z$ and $\theta$. It is also to be noted that for hyperscaling violating backgrounds, $v_B$ is related to the Hawking temperature (temperature of the dual thermal field theory) \cite{Mezei:2016wfz,Blake:2016wvh}.

\section{Holographic Subregion Complexity}\label{sec5}
The Quantum complexity (QC) can be realised in the following way. Considering a simple (unentangled) product state $\ket{\uparrow\uparrow...\uparrow}$ as a reference state, QC is defined as the minimum number of $2$-qubit unitary operation required to prepare a target state $\ket{\psi}$ from the reference state. In this section we study the holographic subregion complexity (HSC) proposal \cite{Alishahiha:2015rta}. This conjecture states that the volume enclosed by the co-dimension two static minimal surface (RT surface) with the boundary coinciding with that of the subsystem, is dual to the complexity of that subregion. For the hyperscaling violating geometry, this co-dimension one volume reads
\begin{eqnarray}\label{24}
V(\Gamma^{min}_A) &=& 2L^{d-2} r_t^{\frac{p+\theta}{p-\theta}-p} \int_{\epsilon/r_t}^{1}du \frac{u^{\frac{2\theta - p}{p-\theta}-p}}{\sqrt{f(u)}}\int_{u}^{1} dk \frac{k^{p+\frac{\theta}{p-\theta}}}{\sqrt{f(k)}\sqrt{1-k^{2p}}}.
\end{eqnarray}
We now use the above volume to obtain the HSC. This is given by (setting AdS radius $R=1$) \cite{Alishahiha:2015rta}
\begin{eqnarray}\label{25}
C_V(A)&=& \frac{V(\Gamma^{min}_A)}{8\pi G_{d+1}}\nonumber\\
&=&\frac{L^{d-2}l}{8\pi G_{d+1} \left(p-\frac{\theta}{p-\theta}\right)\epsilon^{\left(p-\frac{\theta}{p-\theta}\right)}}+\frac{L^{d-2}r_t^{\frac{p+\theta}{p-\theta}-p}}{8\pi G_{d+1}} \sum_{n=0}^{\infty}\sum_{m=0}^{\infty}\bar{V} \left(\frac{r_t}{r_h}\right)^{(m+n)p\left(1+\frac{z}{p-\theta}\right)}
\end{eqnarray}
where 
\begin{eqnarray}
\bar{V}= 	\frac{\Gamma(n+\frac{1}{2})\Gamma(m+\frac{1}{2})\Gamma\left(\frac{(m+n)p(1+\frac{z}{p-\theta})+1+\frac{2\theta}{p-\theta}}{2p}\right)}{\sqrt{\pi}\Gamma(n+1)\Gamma(m+1)\Gamma\left(\frac{(m+n)p(1+\frac{z}{p-\theta})+1+p+\frac{2\theta}{p-\theta}}{2p}\right)}\left[\frac{1}{p\left(mp(1+\frac{z}{p-\theta})+\frac{\theta}{o-\theta}-p\right)}\right]\nonumber~.
\end{eqnarray}
The first term in eq.(\ref{25}) is the divergent piece of the HSC. In the subsequent analysis we will denote it as $C^{div}$. We now consider the small temperature limit, that is $\frac{r_t}{r_h}\ll 1$. In the spirit of this limit, we keep terms up to $\mathcal{O}(m)$ in eq.(\ref{25}). This in turn means that we are interested only in the leading order temperature corrections to the HSC. These considerations lead to the following expression of HSC
\begin{eqnarray}
	C_V(A) = C^{div}+ \frac{L^{d-2}}{8\pi G_{d+1}} \bar{V}_{(0)}r_t^{\frac{p+\theta}{p-\theta}-p} + \frac{L^{d-2}}{8\pi G_{d+1}} \left[\bar{V}_{(1)}+\bar{V}_{(2)}\right]r_t^{\frac{p+\theta}{p-\theta}-p} \left(\frac{r_t}{r_h}\right)^{p\left(1+\frac{z}{p-\theta}\right)}
\end{eqnarray}
where $\bar{V}_{(0)} = \bar{V}|_{(n=0,m=0)}$, $\bar{V}_{(1)} = \bar{V}|_{(n=1,m=0)}$ and $\bar{V}_{(2)} = \bar{V}|_{(n=0,m=1)}$. We now use eq.(\ref{19}) in order to express the above expression in terms of the subsystem size $l$. This is obtained to be
\begin{eqnarray}\label{39}
C_V(A) = C^{div}+C_1l^{(\frac{p+\theta}{p})-(p-\theta)} + C_2 l^{(\frac{p+\theta}{p})+z}T^{1+(\frac{p-\theta}{z})}
\end{eqnarray}
where the expressions for $C_1$ and $C_2$ are given in the Appendix. As we have mentioned earlier, $C^{div}$ represents the divergent piece of HSC whereas the second term is the temperature independent term. The lowest order temperature correction occurs in the third term. This result for the HSC will be used in the next section to compute the mutual complexity between two subsystems.
\section{Complexity for mixed states: Mutual complexity ($\Delta\mathcal{C}$)}\label{sec6} 
The complexity for mixed states (purification complexity) is defined as the minimal (pure state) complexity among all possible purifications of the mixed state. This in turn means that one has to optimize over the circuits which take the reference state to a target state $\ket{\psi_{AB}}$ (a purification of the desired mixed state $\rho_A$) and also need to optimize over the possible purifications of $\rho_A$. This can be expressed as
\begin{eqnarray}
\mathcal{C}(\rho_{A})= \mathrm{min}_{B}~ \mathcal{C}(\ket{\psi_{AB}});~ \rho_A = \mathrm{Tr}_B \ket{\psi_{AB}}\bra{\psi_{AB}}
\end{eqnarray}
where $A^c=B$. Recently a quantity denoted as the `mutual complexity ($\Delta\mathcal{C}$)' has been defined in order to compute the above mentioned mixed state complexity \cite{Alishahiha:2018lfv,Caceres:2018blh}. The computation of $\Delta\mathcal{C}$  starts with a pure state $\rho_{AB}$ in an extended Hilbert space (including auxillary degrees of freedom), then by tracing out the degrees of freedom of $B$, one gets the mixed state $\rho_{A}$. On the other hand, tracing out the degrees of freedom of $A$ yields $\rho_B$. These computed results then can then be used in the following formula to compute the mutual complexity $\Delta\mathcal{C}$ \cite{Alishahiha:2018lfv}
\begin{eqnarray}\label{26}
\Delta\mathcal{C} = \mathcal{C}(\rho_{A})+ +\mathcal{C}(\rho_{B})-\mathcal{C}(\rho_{A\cup B})~.
\end{eqnarray}
The mutual complexity $\Delta\mathcal{C}$ is said to be subadditive if $\Delta\mathcal{C}>0$ and superadditive if $\Delta\mathcal{C}<0$.\\
We now choose to follow the subregion `Complexity=Volume' conjecture to compute the quantities $\mathcal{C}(\rho_{A})$,  $\mathcal{C}(\rho_{B})$ and $\mathcal{C}(\rho_{A \cup B})$. Similary one can follow the `Complexity=Action' conjecture or $C=V2.0$ conjecture \cite{Couch:2016exn} to compute these quantities. We consider two different set up to probe the mutual complexity $\Delta\mathcal{C}$. In the first scenario, we consider two disjoint subsystems $A$ and $B$ of width $l$ on the boundary Cauchy slice $\sigma$. These two subsystems are separated by a distance $x$. We then compute the mutual complexity between these two subregions. Next we consider that the boundary Cauchy slice $\sigma$ is a collection of two adjacent subsystems $A$ and $B$ of width $l$ with $A \cap B =0$ (zero overlap) and $A^c=B$. In this set up we compute the mutual complexity between a subregion $A$ and the full system $A \cup A^c$.
\subsection{Case 1: Mutual complexity between two disjoint subregions}
In this set up, we assume the two subsystems $A$ and $B$ of width $l$ on the Cauchy slice $\sigma$. The separation length between $A$ and $B$ is $x$. We want to see how the rate of complexification of these two subsystems get affected when we introduce correlation (classical and quantum) between these two subregions. In \cite{Auzzi:2019vyh}, the authors have used the `Complexity=Action' conjecture to study the mutual complexity between two subsystems and in \cite{Ghodrati:2019hnn} `Complexity = Volume' conjecture was incorporated to probe $\Delta\mathcal{C}$ between two subregions. Here we follow the approach given in \cite{Ghodrati:2019hnn}.

 \subsubsection{BTZ black hole}
We first compute the HSC corresponding to a single subsystem $A$ of length $l$ for the BTZ black hole. The mentioned black hole geometry is characterized by the following metric \cite{Banados:1992wn,Saha:2018jjb}
  \begin{eqnarray}
  ds^2 = \frac{R^2}{z^2}\left[-f(z) dt^2+\frac{dz^2}{f(z)}+dx^2\right];~f(z)=1-\frac{z^2}{z_h^2}~.
  \end{eqnarray}
Following the prescription of $C=V$ conjecture, the HSC corresponding to a single subsystem $A$ of length $l$ in the dual field theory is obtained by computing the co-dimension $1$ volume enclosed by the co-dimension two RT-surface. This leads to the following
\begin{eqnarray}
\mathcal{C}(\rho_{A})&=& \frac{2R}{8\pi R G_{2+1}} \int_{0}^{z_t} \frac{x(z)}{z^2 \sqrt{f(z)}} dz\nonumber\\
&=& \frac{2R}{8\pi R G_{2+1}} \int_{0}^{z_t} \frac{1}{z^2 \sqrt{f(z)}} dz \int_{z}^{z_t} \frac{du}{\sqrt{f(u)}\sqrt{(\frac{u}{z_t})^2-1}}\nonumber\\
&=& \frac{1}{8\pi} \left(\frac{l}{\epsilon}-\pi\right)~(setting~ G_{2+1}=1)
\end{eqnarray} 
where $\epsilon$ is the cut-off introduced to prevent the UV divergence and $z_t$ is the turning point of the RT surface. It is to be observed that the computed result of HSC in this case is independent of the black hole parameter (that is, event horizon $z_h$). This is a unique feature of AdS$_3$ as the subregion complexity in this case is topological. We now consider two subsystems $A$ and $B$ of equal length $l$, separated by a distance $x$. In this set up, the connected RT surface is governed by two strips of length $2l+x$ and length $x$ \cite{Ben-Ami:2016qex}. This leads to the following \cite{Ghodrati:2019hnn,Ben-Ami:2016qex}
\begin{eqnarray}\label{38}
\mathcal{C}(A \cup B) = \mathcal{C}(2l+x) - \mathcal{C}(x)~.
\end{eqnarray}
Note that when the separation $x$ between the two subsystems vanishes, then
\begin{eqnarray}
\mathcal{C}(l\cup l)=\mathcal{C}(2l)~.
\end{eqnarray}
By substituting eq.(\ref{38}) in eq.(\ref{26}), the mutual complexity is obtained to be
\begin{eqnarray}
\Delta\mathcal{C}
&=& 2 \mathcal{C}(l) - \mathcal{C}(2l+x) + \mathcal{C}(x)= - \frac{1}{4}~.
\end{eqnarray}  
It can be observed that the mutual complexity is less than zero. This implies that the complexity is superadditive. In the above computation of $\mathcal{C}(A\cup B)$, we have considered only the connected RT surface. This is true as long as we work within the limit $\frac{x}{l}\ll 1$. However, there can be disconnected configuration also in which $\mathcal{C}(A\cup B) = \mathcal{C}(A) + \mathcal{C}(B)$ \cite{Ben-Ami:2016qex}. This in turn means that when the separation length $x$ is large enough, mutual complexity between two subregions $\Delta\mathcal{C}$ is zero. This is similar to mutual information.
  
 \subsubsection{Hyperscaling violating geometry}
  By considering the same set up, we now compute the mutual complexity between $A$ and $B$ for the hyperscaling violating geometry. The HSC corresponding to a single strip of length $l$ is given in eq.(\ref{25}). We use this result to compute the complexities $\mathcal{C}(l)$, $\mathcal{C}(2l+x)$ and $\mathcal{C}(x)$. This leads to the following expression of mutual complexity
  \begin{eqnarray}
  \Delta\mathcal{C}&=&2 \mathcal{C}(l) - \mathcal{C}(2l+x) + \mathcal{C}(x)\nonumber\\
  &=& \frac{L^{d-2}}{8\pi G_{d+1}} \sum_{n=0}^{\infty}\sum_{m=0}^{\infty}\bar{V}\Big[2r_t(l)^{\frac{p+\theta}{p-\theta}-p} \left(\frac{r_t(l)}{r_h}\right)^{(m+n)p\left(1+\frac{z}{p-\theta}\right)}+r_t(x)^{\frac{p+\theta}{p-\theta}-p} \left(\frac{r_t(x)}{r_h}\right)^{(m+n)p\left(1+\frac{z}{p-\theta}\right)}\nonumber\\
  &&-r_t(2l+x)^{\frac{p+\theta}{p-\theta}-p} \left(\frac{r_t(2l+x)}{r_h}\right)^{(m+n)p\left(1+\frac{z}{p-\theta}\right)}\Big]
  \end{eqnarray}
  where $r_t(l)$, $r_t(2l+x)$ and $r_t(x)$ corresponds to the turning points of the RT surfaces associated to $l$, $2l+x$ and $x$ respectively. It is to be observed that the divergent pieces of HSC cancels out which yields a finite result. We now consider the small temperature limit $\frac{r_t(x)}{r_h}\ll\frac{r_t(l)}{r_h}\ll\frac{r_t(2l+x)}{r_h}\ll1$ and keep terms up to $\mathcal{O}(m)$. This in turn yields the following expression for mutual complexity
  \begin{eqnarray}
  \Delta\mathcal{C}&=&C_1\left[2l^{(\frac{p+\theta}{p})-(p-\theta)} -(2l+x)^{(\frac{p+\theta}{p})-(p-\theta)}+x^{(\frac{p+\theta}{p})-(p-\theta)}\right]\nonumber\\
  &+&C_2\left[2l^{z+\frac{p+\theta}{p}} -(2l+x)^{z+\frac{p+\theta}{p}}+x^{z+\frac{p+\theta}{p}}\right]T^{1+\left(\frac{p-\theta}{z}\right)}~.
  \end{eqnarray}
Similar to the study of mutual information, we can also point out a critical separation length $x_c$ at which the above expression is zero.
  \begin{figure}[!h]
  	\begin{minipage}[t]{0.48\textwidth}
  		\centering\includegraphics[width=\textwidth]{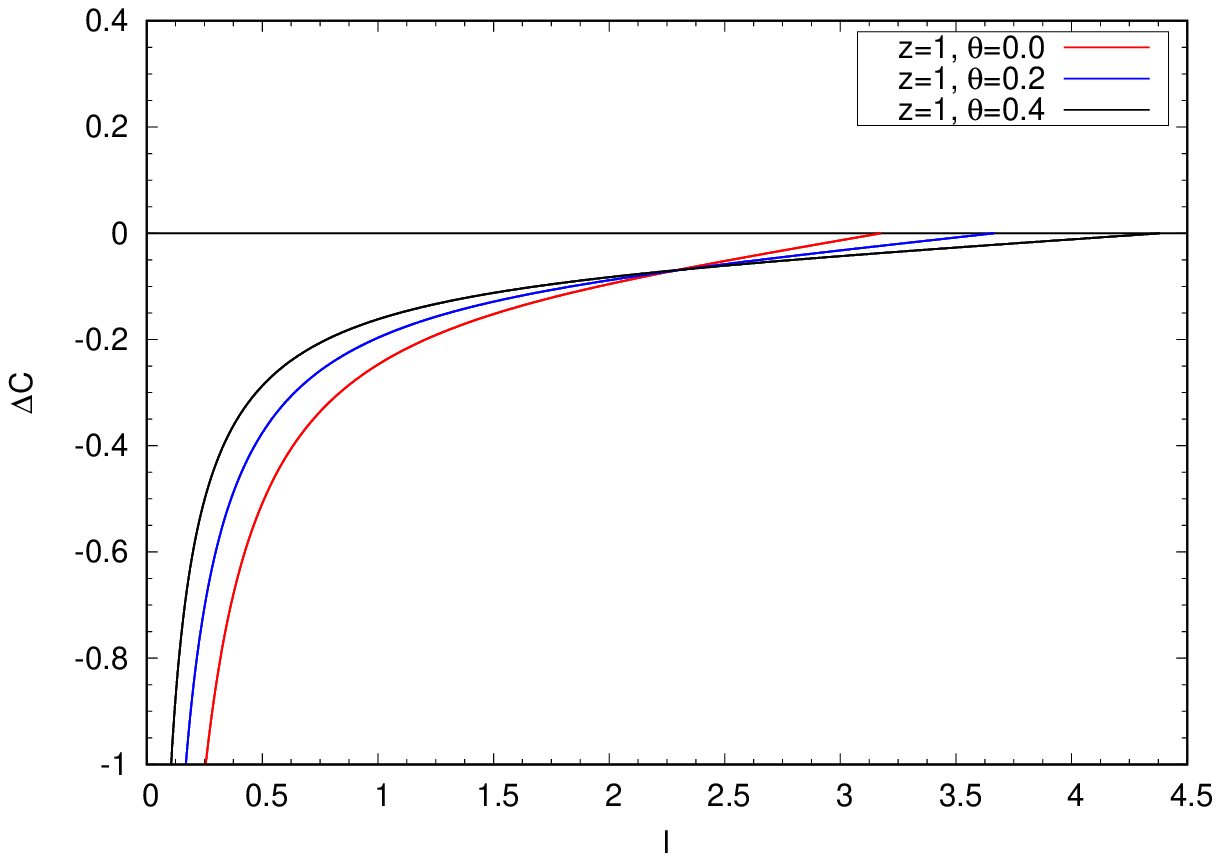}\\
  		{\footnotesize Effect of hyperscaling violating exponent $\theta$ (we set $z=1$)}
  	\end{minipage}\hfill
  	\begin{minipage}[t]{0.48\textwidth}
  		\centering\includegraphics[width=\textwidth]{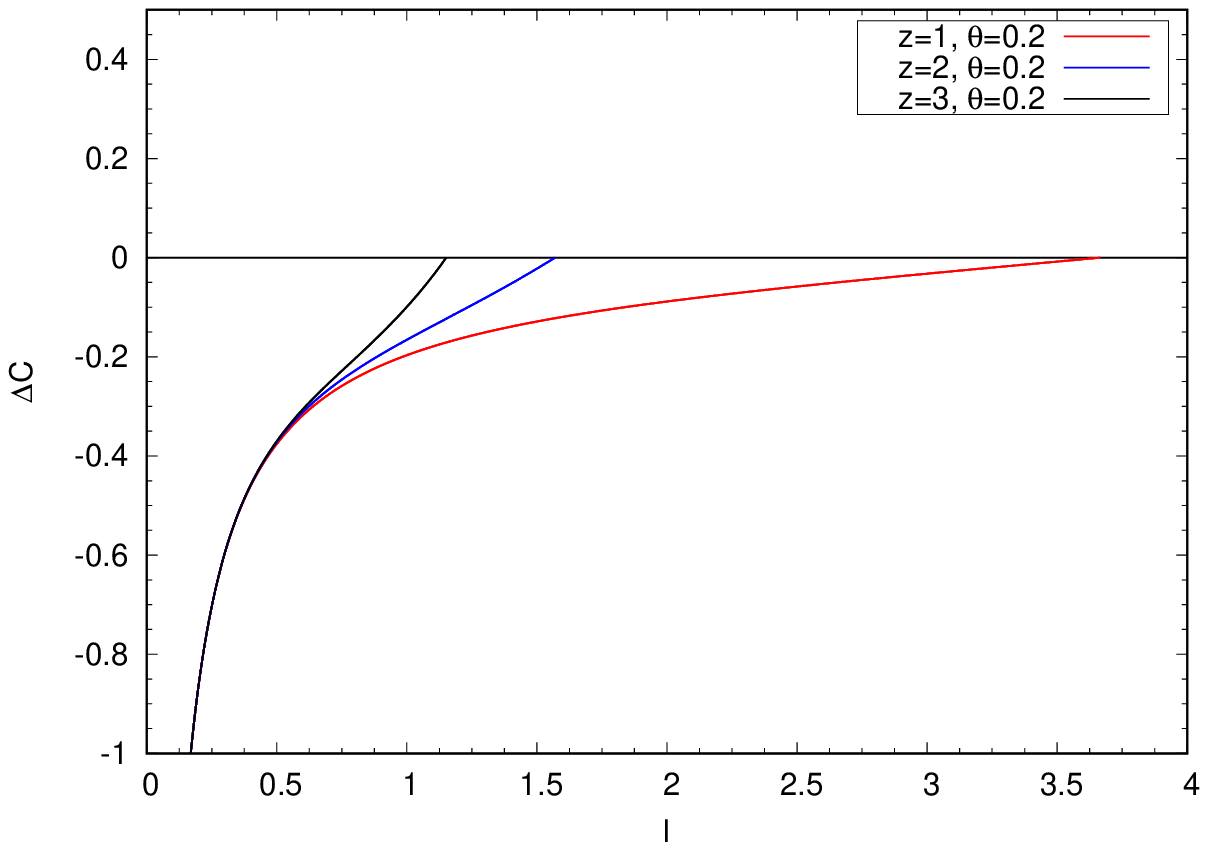}\\
  		{\footnotesize Effect of dynamical exponent $z$ (we set $\theta=0.2$)}
  	\end{minipage}
  	\caption{Effects of $\theta$ and $z$ on $\Delta\mathcal{C}$  (with $d=3$, $k=0.4$, $L=1$ and $G_{d+1}=1$)}
  	\label{fig4}
  \end{figure} 
In Fig.(\ref{fig4}), we have graphically represented the computed result of $\Delta\mathcal{C}$. In the above plots, we have introduced $k=\frac{x}{l}$ in order to compute the critical separation length $x_c$ at which $\Delta\mathcal{C}=0$. These plots also represent the collective effects of $z$ and $\theta$ on the mutual complexity.
\subsection{Case 2: Mutual complexity between two adjacent subsystems}
We now consider that the boundary Cauchy slice $\sigma$ is composed of two adjacent subsystems $A$ and $B$ of width $l$. Further, we assume $A \cap B =0$ (zero overlapping) and $A^c=B$ and the full system (on $\sigma$) is in a pure state. In this set up we compute the mutual complexity between a subregion $A$ and the full system $A \cup A^c$ \cite{Agon:2018zso,Caceres:2019pgf}.
\subsubsection{BTZ black hole}
We now proceed to compute the mutual complexity between $A$ and $B=A^c$. In this set up, the connected RT surface is composed of one strip of length $2l$. This leads to the following expression for mutual complexity
\begin{eqnarray}
\Delta \mathcal{C} &=& 2\mathcal{C}(l)-\mathcal{C}(2l)\nonumber\\
 &=&\frac{1}{8 \pi} \left[\frac{l}{\epsilon}-\pi+\frac{l}{\epsilon}-\pi-\frac{2l}{\epsilon}+\pi\right]\nonumber\\
  &=& - \frac{1}{8}.
\end{eqnarray}
Similar to the disjoint subregion case, mutual complexity in this set up is also superadditive. This in turn means that the complexity of the state corresponding to the full system is greater than the sum of the complexities of the states in the two subsystems.
\subsection{Hyperscaling violating geometry}
We now proceed to compute the mixed state complexity for the hyperscaling violating geometry. By using the HSC result given in eq.(\ref{25}), $\Delta\mathcal{C}$ in this set up reads
\begin{eqnarray}\label{27}
\Delta\mathcal{C} &=& \frac{1}{8\pi G_{d+1}} \left[V(\Gamma^{min}_A)+V(\Gamma^{min}_B)-V(\Gamma^{min}_{A \cup B})\right]\nonumber\\
&=&\frac{L^{d-2}}{8\pi G_{d+1}}\sum_{n=0}^{\infty}\sum_{m=0}^{\infty}\bar{V}\left[2r_t(l)^{\left(\frac{p+\theta}{p-\theta}\right)-p}\left(\frac{r_t(l)}{r_h}\right)^{(m+n)p\left(1+\frac{z}{p-\theta}\right)}-r_t(2l)^{\left(\frac{p+\theta}{p-\theta}\right)-p}\left(\frac{r_t(2l)}{r_h}\right)^{(m+n)p\left(1+\frac{z}{p-\theta}\right)}\right]~.\nonumber\\
\end{eqnarray}  
We now consider the limit $\frac{r_t(l)}{r_h}\ll \frac{r_t(2l)}{r_h} \ll 1 $. In this limit, we keep terms up to order $\mathcal{O}(m)$ in the above expression. This in turn leads to the following expression
\begin{eqnarray}\label{28}
\Delta\mathcal{C} =  \left[2-2^{(\frac{p+\theta}{p})-(p-\theta)}\right]C_1l^{\left(\frac{p+\theta}{p}\right)-\left(p-\theta\right)}+\left[2-2^{z+\left(\frac{p+\theta}{p}\right)}\right]C_2 l^{z+\left(\frac{p+\theta}{p}\right)}T^{1+\left(\frac{p-\theta}{z}\right)}~.
\end{eqnarray}
 We observe that similar to the BTZ case, the above result for $\Delta C$ is less than zero. This in turn means that the mutual complexity computed using the HSC conjecture yields a superadditive result.

\section{Conclusion}\label{sec7}
In this paper, we compute the entanglement entropy and complexity for mixed states by using the gauge/gravity correspondence. We start our analysis by considering a hyperscaling violating solution as the bulk theory. This geometry is associated with two parameters, namely, hyperscaling violating exponent $z$ and dynamical exponent $\theta$. It is dual to a non-relativistic, strongly coupled theory with hidden Fermi surfaces. We then consider a single strip-like subsystem in order to compute the HEE of this gravitational solution. We observe that the computed result of HEE along with the internal energy $E$, satisfies a Smarr-like thermodynamics relation associated with a generalized temeperature $T_g$. This thermodynamic relation naturally emerges by demanding that the generalized temperature $T_g$ reproduces the Hawking temperature $T_H$ as the leading term in the IR ($r_t \rightarrow r_h$) limit. In UV limit ($\frac{r_t}{r_h}\ll 1$), it is found that $T_g \propto \frac{1}{l^z}$, that is, $T_g$ is inversely proportional to subsystem size $l$. This behaviour is compatible with the definition of entanglement temperature given in the literature. We then holographically compute the relative entropy $S_{rel}$, by incorporating the perturbative approach. Using this the Fisher information metric is computed. We find that in this case the power of $l$ carries both the exponents $z$ and $\theta$. We then consider two strip-like subsystems $A$ and $B$ separated by a length $D$, in order to compute the EWCS ($E_W$) which is the holographic analogy of EoP. We compute $E_W$ for both low and high temperature conditions. In both cases, there is a temperature independent term (denoted as $E_W^{T=0}$) which is independent of the hyperscaling violating exponent $z$ but depends on the dynamical exponent $\theta$. On the other hand for a large enough value of $D$ (critical separation length $D_c$), the RT surface $\Gamma_{AB}^{min}$ becomes disconnected and $E_W$ should vanish. This in turn means that $E_W$ probes the phase transition between the connected and disconnected phases of the RT surface $\Gamma_{AB}^{min}$. We evaluate these critical separation point $D_c$ by using the property that at $D_c$ the mutual information between $A$ and $B$ becomes zero as they become disconnected. This behaviour for $I(A:B)$ and $E_W$ is shown in Fig(s).(\ref{fig2}, \ref{fig3}) for both low and high temperature cases. We observe that $E_W$ always satisfies the property $E_W > \frac{1}{2} I(A:B)$. We then discuss the property of information spreading by computing the Butterfly velocity. We observe that the computed expression of Butterfly velocity explicitly depends on the temperature of the dual field theory. We then compute the HSC by considering again a single strip-like subsystem. The complexity for mixed state is computed by following the concept of mutual complexity $\Delta C$. We have used the HSC conjecture to compute the $\Delta C$ for both BTZ black hole and hyperscaling violating geometry. We have studied the mutual complexity by considering two different set ups. Firstly, we consider two disjoint subsystems $A$ and $B$ of width $l$, separated by a length $x$ on the boundary Cauchy slice $\sigma$. Computation of $\Delta\mathcal{C}$ in this set up probes rate of complexification of these two subsystems when we consider correlation (both classical and quantum) between them. Next we consider a single subsystem $A$ in such a way that $A^c=B$ and $A \cap B=0$. We then measure the mutual complexity between a subsystem $A$ and the full system $A\cup A^c$. We observe that the computed result of mutual complexity is superadditive that is $\Delta C <0$. This in turn means that the complexity of the state corresponding to the full system is greater than the sum of the complexities of the states in the two subsystems. We observe that for BTZ black hole $\Delta \mathcal{C}$ is independent of temeprature however for hyperscaling violating solution it contains a temperature independent term as well as a temperature dependent term. It is to be kept in mind that this nature of $\Delta C$ is observed for the HSC conjecture and similarly one can use `Complexity=Action' conjecture \cite{Alishahiha:2018lfv,Auzzi:2019vyh}  or `CV2.0' conjecture to compute $\Delta C$ in this context \cite{Agon:2018zso}.
\section*{Acknowledgements}
A.S. would like to acknowledge the support by Council of Scientific and Industrial Research (CSIR, Govt. of India) for Senior Research Fellowship.
\section*{Appendix}
In this appendix, we give the expressions of quantities that appear in the main text. These are as follows.
\begin{eqnarray*}
\Delta_4&=& \Delta_3 \left(\frac{p}{\sqrt{\pi}}\frac{\Gamma\left(\frac{2p+\frac{p}{p-\theta}}{2p}\right)}{\Gamma\left(\frac{p+\frac{p}{p-\theta}}{2p}\right)}\right)^{(p-\theta)\left(\frac{1-p}{p}\right)}-\frac{\left(\frac{4\pi}{p(1+\frac{z}{p-\theta})}\right)^{1+\frac{p-\theta}{z}}}{2\left(p(1+\frac{z}{p-\theta})-p+1\right)}\left(\frac{p}{\sqrt{\pi}}\frac{\Gamma\left(\frac{2p+\frac{p}{p-\theta}}{2p}\right)}{\Gamma\left(\frac{p+\frac{p}{p-\theta}}{2p}\right)}\right)^{\left(\frac{p-\theta}{p}\right)\left(1+\frac{pz}{p-\theta}\right)}\nonumber\\
\Delta_5 &=& \frac{\sqrt{\pi}}{(p-1)}\left(\frac{4\pi}{p(1+\frac{z}{p-\theta})}\right)^{(p-1)(\frac{p-\theta}{pz})}\frac{\Gamma\left(\frac{1+\frac{pz}{p-\theta}}{p(1+\frac{z}{p-\theta})}\right)}{\Gamma\left(\frac{2-p+\frac{pz}{p-\theta}}{2p(1+\frac{z}{p-\theta})}\right)}\nonumber\\
\end{eqnarray*}
\begin{eqnarray*}
C_1 &=& \frac{L^{d-2}}{8\pi G_{d+1}} \bar{V}_{(0)} \left(\frac{p}{\sqrt{\pi}}\frac{\Gamma\left(\frac{2p+\frac{p}{p-\theta}}{2p}\right)}{\Gamma\left(\frac{p+\frac{p}{p-\theta}}{2p}\right)}\right)^{(\frac{p+\theta}{p})-(p-\theta)}\nonumber\\
C_2 &=& \left[(\bar{V}_{(1)}+\bar{V}_{(2)})\left(\frac{p}{\sqrt{\pi}}\frac{\Gamma\left(\frac{2p+\frac{p}{p-\theta}}{2p}\right)}{\Gamma\left(\frac{p+\frac{p}{p-\theta}}{2p}\right)}\right)^{(\frac{p+\theta}{p})+z}-\left(\frac{p+\theta}{p-\theta}-p\right)\bar{V}_{(0)}\Delta_3\left(\frac{p}{\sqrt{\pi}}\frac{\Gamma\left(\frac{2p+\frac{p}{p-\theta}}{2p}\right)}{\Gamma\left(\frac{p+\frac{p}{p-\theta}}{2p}\right)}\right)^{(\frac{p+\theta}{p})-(p-\theta)}\right]\nonumber\\
&&\times \frac{L^{d-2}}{8\pi G_{d+1}} \left[\frac{4\pi}{p(1+\frac{z}{p-\theta})}\right]^{1+(\frac{p-\theta}{z})}~.\nonumber
\end{eqnarray*}
\bibliographystyle{hephys}  
\bibliography{References}
\end{document}